\documentstyle[aps,version2,preprint,epsfig,rotating]{revtex}
\begin{document}
\draft
\preprint{OCIP/C 97-02}
\preprint{hep-ph/9703285}
\preprint{March 1997}
\begin{title}
What Can We Learn About Leptoquarks At LEP200?
\end{title}
\author{Michael A. Doncheski$^1$ and Stephen Godfrey$^2$}
\begin{instit}
$^1$Department of Physics, Pennsylvania State University, Mont Alto, PA 
17237 USA \\
$^2$Ottawa-Carleton Institute for Physics \\
Department of Physics, Carleton University, Ottawa CANADA, K1S 5B6
\end{instit}

\begin{abstract}
We investigate the  discovery potential for first generation
leptoquarks at the LEP200 $e^+e^-$ collider.  
We consider direct leptoquark searches using 
single leptoquark production via 
resolved photon contributions which offers a much higher kinematic 
limit than the more commonly considered leptoquark pair production 
process.  Depending on the coupling strength of the leptoquark, search 
limits can be obtained to within a few GeV of $\sqrt{s}$.
We also consider LQ limits that can be obtained from t-channel 
interferences effects in $e^+e^- \to hadrons$.

\end{abstract}

With the recent observation of an excess of high $Q^2$ events in $ep$ 
collisions by the H1 \cite{h197} and ZEUS \cite{zeus97} collaborations and 
the possibility that these events signal the existence of leptoquarks
--- colour
(anti-)triplet, spin 0 or 1 particles which carry both baryon and lepton 
quantum numbers --- there is considerable interest in the study of 
these particles.
Leptoquarks appear in a large number of extensions 
of the standard model such as grand unified theories, technicolour, 
and composite models.  The signature for leptoquarks is very
striking: a high $p_{_T}$ lepton balanced by a jet (or missing $p_{_T}$ 
balanced by a jet, for the $\nu q$ decay mode, if applicable). 
Previous 
searches for leptoquarks have been performed by the H1 \cite{H1} and 
ZEUS \cite{ZEUS}
collaborations at the HERA $ep$ collider, by the D0 \cite{D0} 
and CDF \cite{CDF} 
collaborations at the Tevatron $p\bar{p}$ collider, and by the ALEPH 
\cite{ALEPH}, 
DELPHI \cite{DELPHI}, L3 \cite{L3}, and OPAL\cite{OPAL} 
collaborations at the LEP $e^+e^-$ collider.  

In this communication we examine the information that can obtained 
about leptoquarks from $e^+ e^-$ collisions at the LEP200 $e^+e^-$ 
collider at CERN.  Information can be obtained primarily using three 
different approaches. In the first, LQ limits are obtained from LQ 
pair production \cite{h-r}.  Limits can be obtained up to essentially 
$M_{LQ} \sim\sqrt{s}/2$.  These limits have been surpassed considerably 
by the limits obtained at HERA and the Tevatron so we will 
not mention this approach again.
The second approach is single 
leptoquark production in $e^+e^-$ collisions which utilizes the quark 
content of a Weizacker-Williams photon radiating off of one of the 
initial leptons\cite{DG3,DG1,DG2,eboli,leptoLN,HP,misc}.  
This process offers the advantage of a much higher 
kinematic limit than the LQ pair production process, is independent of 
the chirality of the LQ, and gives similar results for both scalar and 
vector leptoquarks.   We will 
concentrate on the limits that can be obtained from this approach.  
The final approach is to search for deviations from standard model 
predictions for the
$e^+e^-\to q\bar{q} \to hadrons$ cross section which might arise from 
t-channel leptoquark exchange \cite{h-r,opal-pn280}.  We find that 
measurements that can be made at LEP200 complements those from HERA 
and the Tevatron.

The  most general $SU(3)\times SU(2) \times U(1)$ invariant scalar 
leptoquarks satisfying baryon and lepton number conservation have 
been written down Buchm\"uller {\sl et al.}\cite{buch}.  However, 
only those leptoquarks which couple to electrons can be produced in 
$e\gamma$ collisions so that we only consider the production of 
leptoquarks coupling to first generation fermions.
Further, for real leptoquark production the chirality 
of the coupling is irrelevant.  For this case the number of 
leptoquarks reduces to four which can be distinquished by their 
electromagnetic charge; $Q_{em}= -1/3$, $-2/3$, $-4/3$, and $-5/3$.  
In our calculations we will sometimes follow the convention
where the leptoquark couplings are replaced 
by a generic Yukawa coupling $g$ which is scaled to electromagnetic 
strength $g^2/4\pi=\kappa \alpha_{em}$ with $\kappa$ allowed to vary.

The process we are considering is shown if Fig. 1.  The parton level 
cross section is trivial, given by: 
\begin{equation}
\sigma(\hat{s})=\frac{\pi^2 \kappa \alpha_em}{M_s} 
                \delta(M_s - \sqrt{\hat{s}})
\end{equation}
for scalar LQ's.  For vector LQ's the cross section is a factor of two 
larger.
Convoluting the parton level cross section with the quark 
distribution in the photon one obtains the expression
\begin{eqnarray}
\sigma(s) & = & \int f_{q/\gamma}(z,M_s^2) \hat{\sigma}(\hat{s}) dz 
                \nonumber \\
& = & f_{q/\gamma}(M_s^2/s,M_s^2) 
      \frac{\mbox{$2\pi^2\kappa \alpha_{em}$}}{\mbox{$s$}}.
\end{eqnarray}
This cross section depends on the LQ charge through $f_{q/\gamma}$ since 
the photon has a larger $u$ quark content than $d$ quark content and 
hence has a larger cross section for LQ's which couple to the $u$ 
quark.  For $e^+e^-$ colliders
the cross section is obtained by 
convoluting the expression for the resolved photon contribution to 
$e \gamma$ production of leptoquarks, Eqn. (2), with the 
Weizs\"acker-Williams effective photon distribution:
\begin{equation}
\sigma(e^+ e^- \rightarrow X S) = \frac{2 \pi^2 \alpha_{em}\kappa}{s} 
    \int_{M_s^2/s}^1 \frac{dx}{x} f_{\gamma/e}(x,\sqrt{s}/2) 
    f_{q/\gamma}(M_s^2/(x s), M_s^2).
\end{equation}

%\begin{figure}[h]
\vskip 0.2cm
\centerline{\epsfig{file=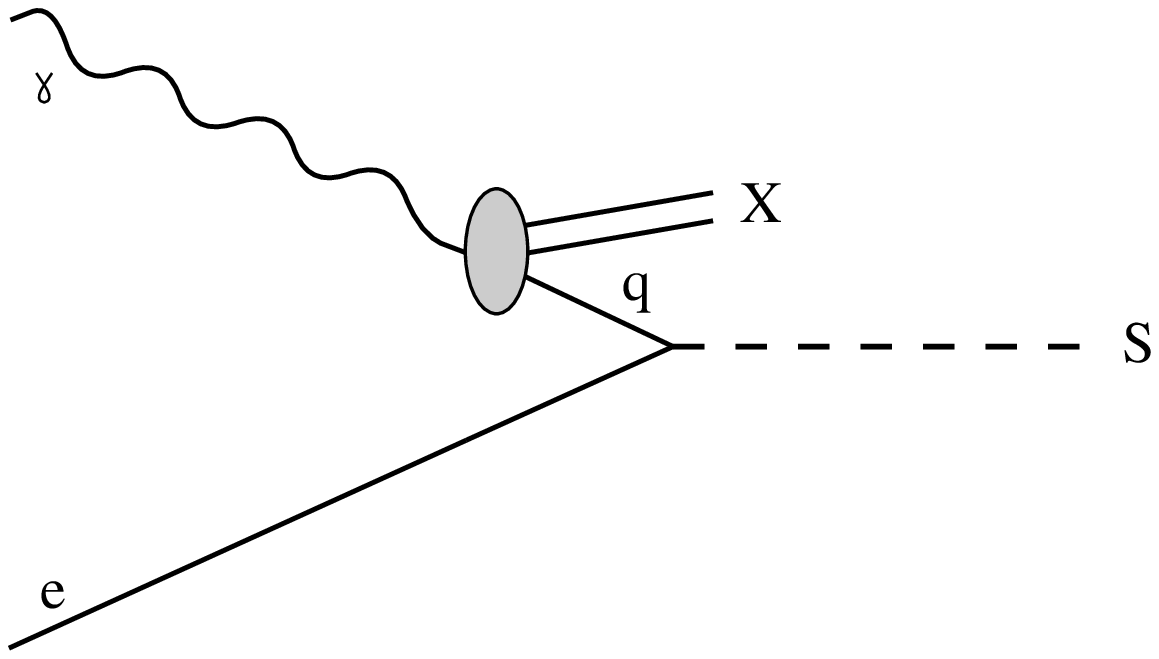,width=5.5cm,clip=}}
\noindent
{\small 
{\bf Fig. 1:} The resolved photon contribution for leptoquark production in 
$e\gamma$ collisions.}
%\end{figure}

There exist several different quark distribution functions in the 
literature \cite{nic,DO,DG,GRV,LAC}.  
The different distributions give almost identical results 
for the $Q_{LQ}=-1/3, \; -5/3$ leptoquarks  
and for the $Q_{LQ}=-2/3, \; -4/3$ leptoquarks 
give LQ cross sections that vary by most a factor 
of two, depending on the kinematic region.  We obtain our results using
the GRV distribution functions \cite{GRV} which we take to be 
representative of the quark distributions in the photon.

We next consider possible backgrounds 
\cite{DGP}.  The leptoquark signal consists of 
a jet and electron with balanced transverse momentum and 
possibly activity from the hadronic remnant of the photon.  
The only serious background is a hard scattering of a quark inside the 
photon by the incident lepton via t-channel photon exchange; $eq \to 
eq$.  We plot the invariant mass distribution for this background in 
our plots of the LQ cross sections and find that it is typically 
smaller than our signal by two orders of magnitude.
For the LQ invariant mass distribution we chose a 5~GeV invariant mass 
bin so that $d\sigma/dM =\sigma/5$~GeV.
Related to this process is the direct production of a 
quark pair via two photon fusion
\begin{equation}
e + \gamma \to e + q  + \bar{q}.
\end{equation}
However, this process is dominated by the collinear divergence which 
is actually well described by the resolved photon process $eq\to eq$ 
given above.  Once this contribution is subtracted away the remainder 
of the cross section is too small to be a concern \cite{DGP}.
Another possible background consists of $\tau$'s pair produced via 
various mechanisms with one $\tau$ decaying leptonically and the other 
decaying hadronically.  Because of the neutrinos in the final state it  
is expected that the electron and jet's $p_T$ do not in general 
balance which would distinguish these backgrounds from the signal.
However,  this background should be checked in a realistic 
detector Monte Carlo to be sure.

In Fig. 2 we show the single LQ production
cross sections for $\sqrt{s}=184$, 190, and 
200~GeV.  In Fig. 3 we use these cross sections to obtain estimates of 
the search limits on scalar leptoquarks that might be achieved 
at LEP200 as a function of mass and Yukawa couplings.
In our results we assume $BR(LQ \to e + q)=1$.  If instead 
$BR(LQ \to e + q)=0.5 $ 

\newpage

\vskip 0.3cm
\begin{minipage}[t]{5.0cm}
\centerline{
\begin{turn}{90}
\epsfig{file=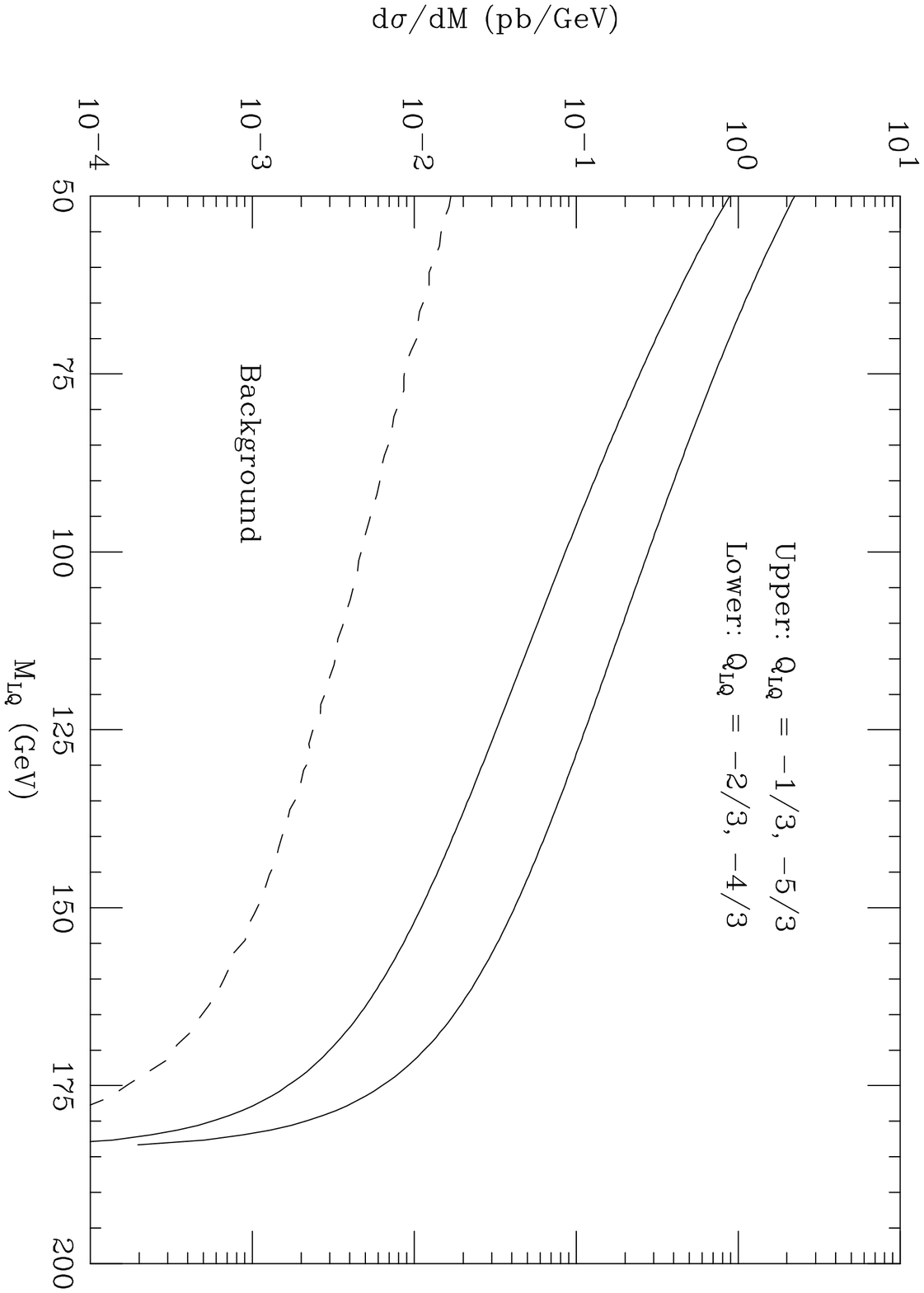,width=3.75cm,clip=}
\end{turn}}
\end{minipage} \
\begin{minipage}[t]{5.0cm}
\centerline{
\begin{turn}{90}
\epsfig{file=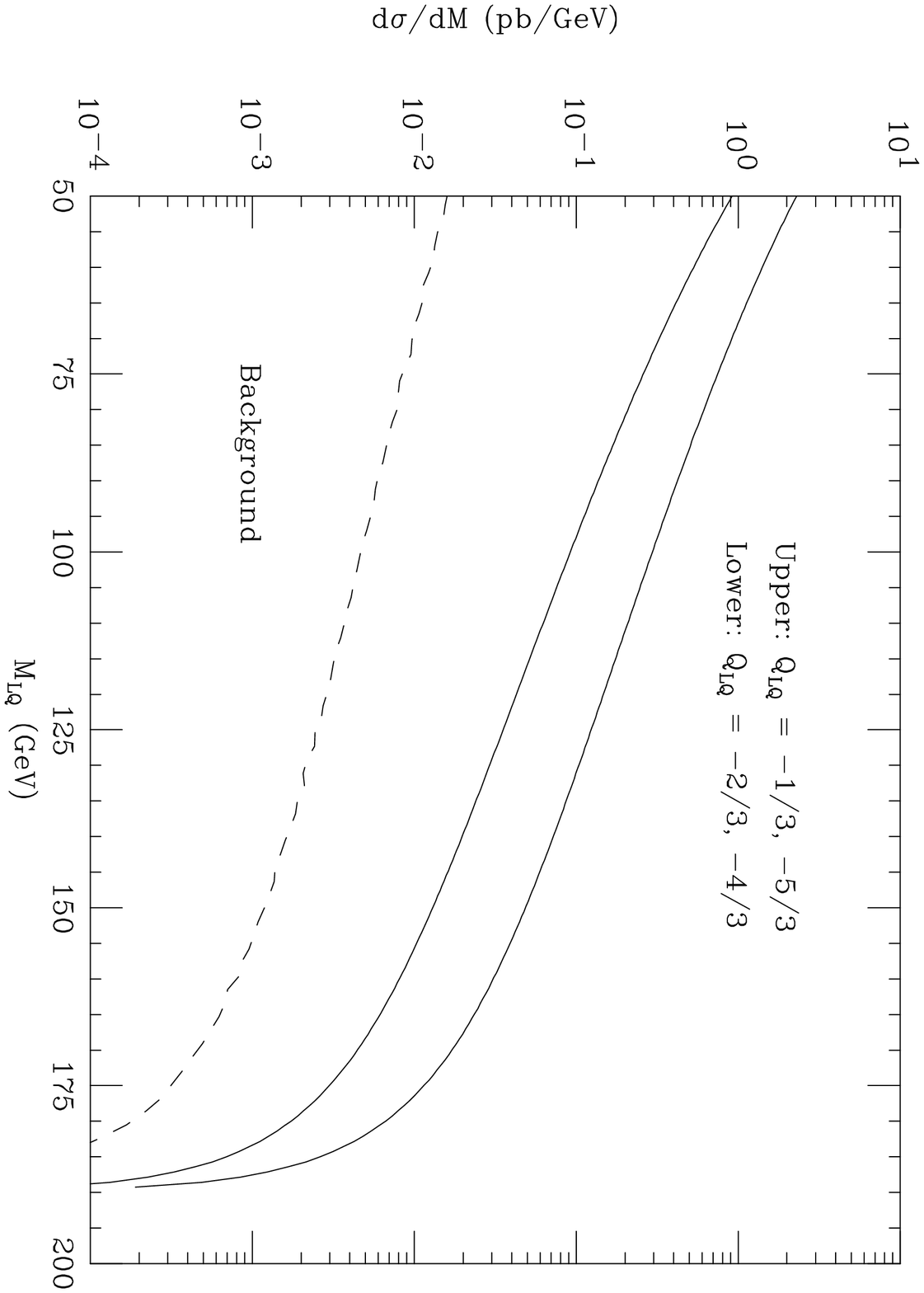,width=3.75cm,clip=}
\end{turn}}
\end{minipage} \
\begin{minipage}[t]{5.0cm}
\centerline{
\begin{turn}{90}
\epsfig{file=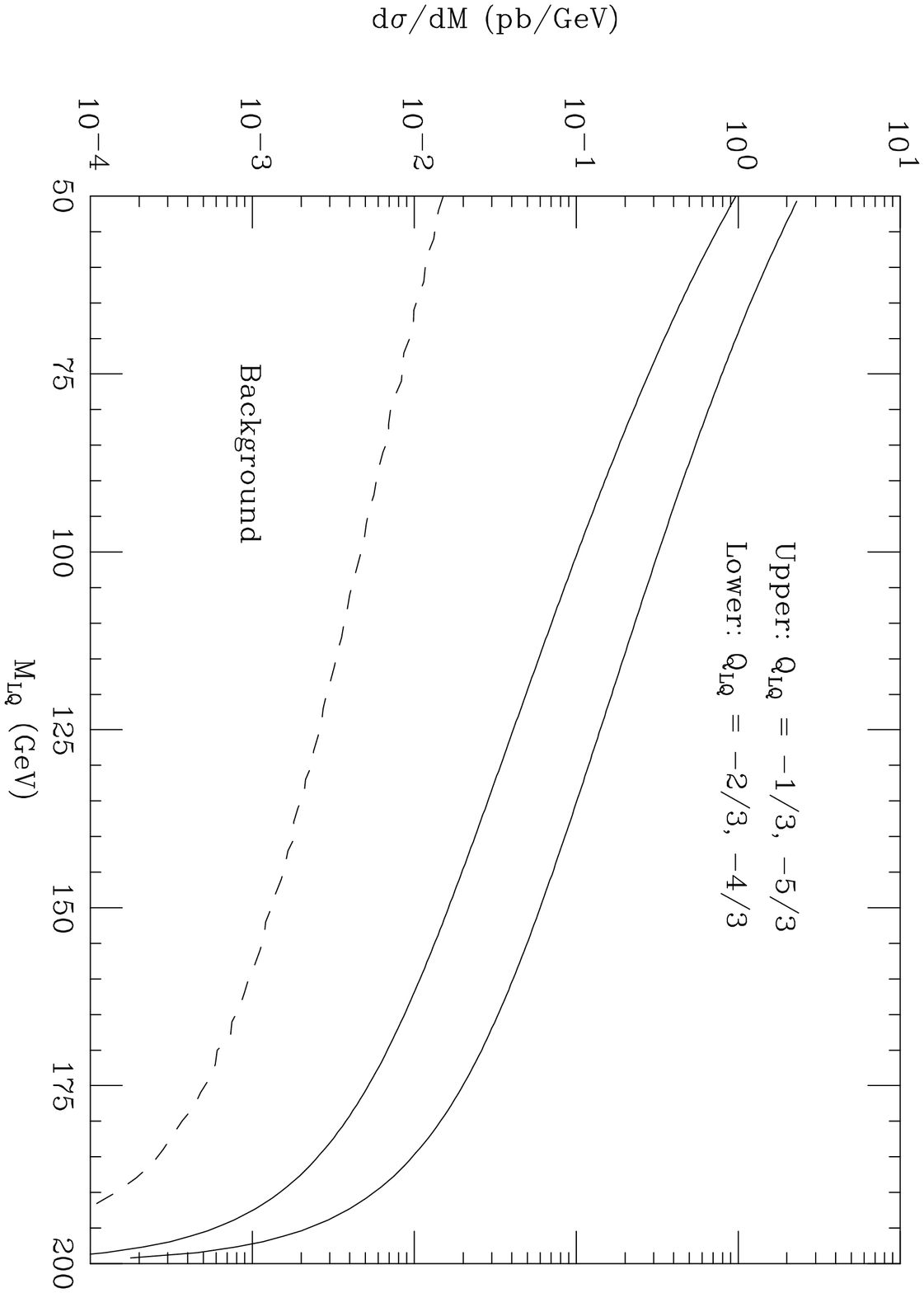,width=3.75cm,clip=}
\end{turn}}
\end{minipage} 

\noindent
{\small {\bf Fig. 2:}
The cross sections for scalar leptoquark production due to resolved 
photon contributions in $e^+ e^-$ collisions for (a) $\sqrt{s}=184$~GeV,
(b) 190~GeV, and (c) 200~GeV.
$\kappa$ is chosen to be 1 and the resolved photon distribution 
functions of Gl\"uck, Reya and Vogt \cite{GRV}
are used. The dashed line is the $e[q]_\gamma \to e q$ background. 
For the LQ invariant mass distribution we use a 5~GeV invariant mass 
bin so that $d\sigma/dM =\sigma/5$~GeV. }

\vskip 0.5cm

\noindent
and $BR(LQ \to \nu + q)=0.5$ the second LQ 
decay mode would have an even more dramatic signature than the one we 
consider; a high $p_T$ monojet balanced against a large missing $p_T$.
Thus,  in this case the sum of the two possible decays would give 
similar limits.  We define our limits as the combination of LQ mass 
and coupling that would result in 10 $e-jet$ events with the correct 
topology for a given integrated luminosity for the four LEP 
experiments combined.  Because we do not know 
for certain what the total integrated luminosity will be at these 
energies we use the following four values of integrated luminosity
to obtain results;
a pessimistic 200~pb$^{-1}$ ($4\times 50$),
an expected (for the 184~GeV run) 400~pb$^{-1}$ ($4\times 100$), 
an expected (for the 190~GeV run) 1000~pb$^{-1}$ ($4\times 250$),
and an optimistic 2000~pb$^{-1}$ ($4\times 500$).
The limits are relatively insensitive to the exact value of the 
luminosity at large values of 
the Yukawa coupling but become fairly sensitive as the strength of the
Yukawa coupling decreases.  Because the vector LQ cross section is 
twice that of the scalar LQ cross section we can obtain the vector LQ 
limits for a given luminosity by using the curves for the next higher 
luminosity for the scalar case. (ie.  the limits for vector LQ's with 
200~pb$^{-1}$ is given by the curve for the scalar case with
400~pb$^{-1}$.)
The limits that can be obtained from single LQ production are
quite competitive with 
limits obtained by the Tevatron experiments \cite{D0,CDF}.  In certain 
regions of the parameter space (small values of the Yukawa coupling)
the limits

\newpage

\centerline{ \epsfig{file=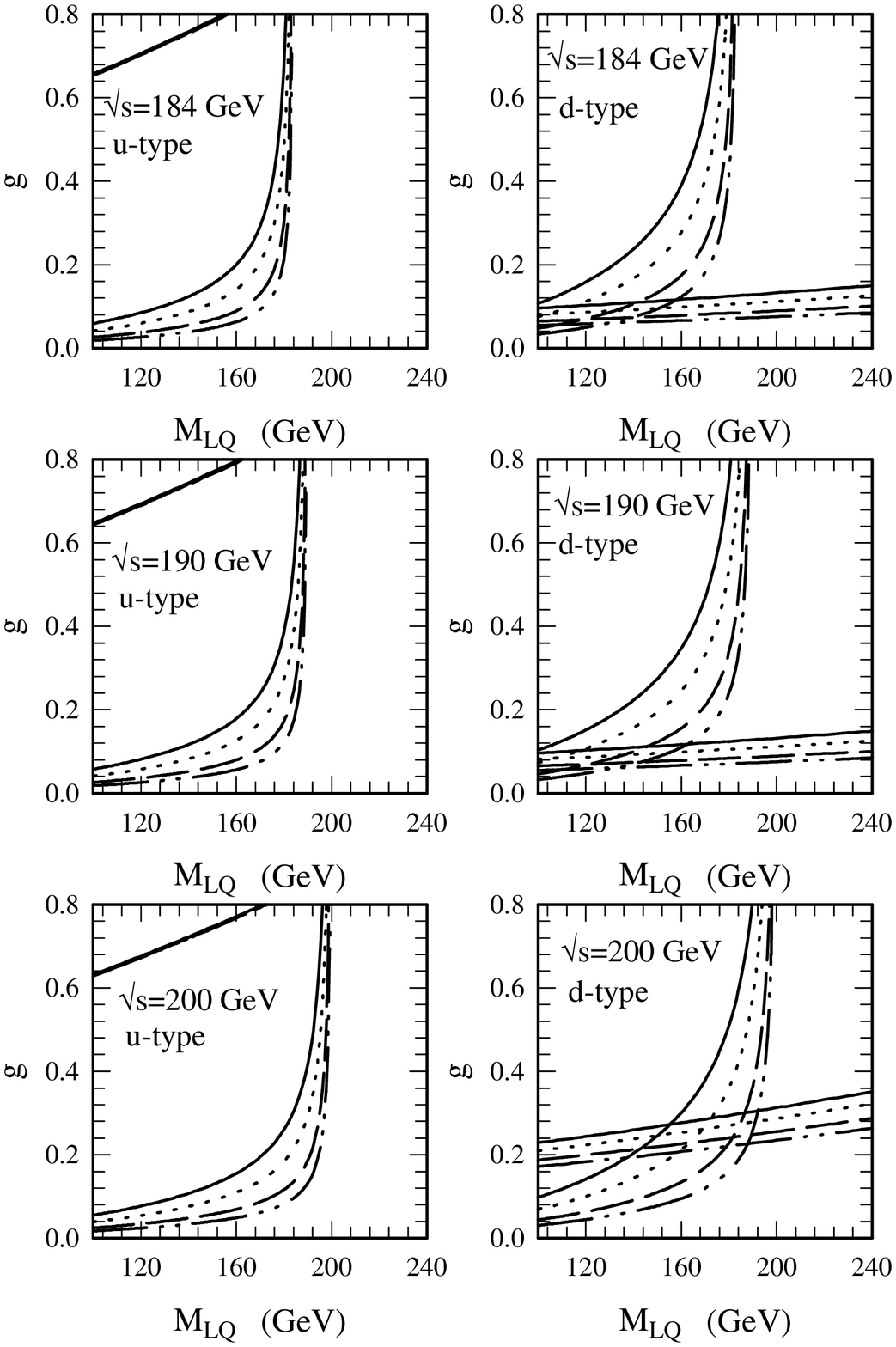,width=9.0cm,clip=} }
\noindent
{\small {\bf Fig. 3:} Exclusion regions for the LQ coupling, $g$
as a function of $M_{LQ}$. 
The region above and to the right of the curved lines would be excluded by
the nonobservation of 
at least 10 single LQ events for a given integrated luminosity. 
The region above the horizontal lines defines the region that 
could be excluded using the 
contributions of t-channel scalar LQ exchange in $e^+e^-\to hadrons$.  
In all cases the solid line is for 
L=200~pb$^{-1}$, the dotted line for L=400~pb$^{-1}$, the dashed line 
is for L=1000~pb$^{-1}$, and the dot-dot-dashed line is for 
L=2000~pb$^{-1}$.} 

\vskip 0.5cm

\noindent
are also competitive with HERA results, in some cases 
they are even more stringent.
We note that these limits are extracted at the kinematic 
limit where $x\to 1$ and the hadronic remnant of the photon has 
vanishing energy.  In this kinematic region the factorization into 
struck parton and remnant is questionable with the quark distribution 
functions subject to higher twist effects.
Nevertheless, despite these qualifications,  we believe 
our estimates to be fairly 
robust and are not likely to be changed substantially by a more rigorous 
scrutinization.

Finally, we comment on the sensitivity of the process $e^+e^- \to hadrons$
to LQ's via t-channel LQ exchange.  In Fig. 3 we 
include limits based on comparing deviations expected from LQ exchange to 
the 1-sigma statistical errors assuming
standard model cross sections.  We include these results primarily to 
remind the reader that precision results can put stringent limits on 
new physics.  For LQ's coupling to u-quarks the limits are rather weak 
but the limits on LQ's coupling to d-quarks the limits are rather 
stringent. Thus, cross section measurements can be sensitive to the 
existence of LQ's up to many times $\sqrt{s}$, depending on the LQ 
coupling. If the recent HERA results are confirmed by better 
statistics LEP200 measurements could play an important role in 
understanding the basis for these anomalies.  Having said this we 
stress that we only wish to draw attention to the fact that these 
measurements are potentially useful.  Our analysis is hopelessly 
naive, not having taken into account experimental acceptances and 
systematic errors.  To further emphasize this, a
recent analysis by the OPAL collaboration \cite{opal-pn280} using measurements 
of $e^+e^- \to hadrons$ taken at 133~GeV, 161~GeV, and 172~GeV and 
employing a one-sided likelihood fit obtains more stringent limits 
than ours for LQ's coupling to the u-quark but weaker limits for LQ's 
coupling to the d-quark.  The difference is due to the fact that the 
experimental measurements are in the wrong direction to the changes 
expected from u-type LQ's but in the right direction for d-type LQ's.

In this communication we have pointed out that information 
about LQ's that can be obtained at LEP200 complements measurements made 
at other colliders such as HERA and the Tevatron.  
We have used the resolved photon contributions to 
single leptoquark production and t-channel leptoquark exchange 
to estimate potential limits on leptoquark masses and couplings.
If the recent HERA results are confirmed, measurements at LEP200 could 
play an important role in undertanding the underlying physics.
Finally, we remind the reader 
that our results are of course only 
theorist's estimates which should be examined more closely and 
carefully than has been described here.

\acknowledgments

This research was supported in part by the Natural Sciences and Engineering 
Research Council of Canada. The authors thank Bob Carnegie and Richard 
Hemingway for helpful conversations.


\begin{references}

\bibitem{h197}
H1 Collaboration, C. Adloff {\it et al.}, DESY Report 97-24 (Feb. 1997).

\bibitem{zeus97}
ZEUS Collaboration, J. Breitweg, {\it et al.}, DESY Report 97-25 (Feb. 1997).

\bibitem{H1}
H1 Collaboration, I. Abt {\it et al.}, Nucl. Phys. {\bf B396}, 3 (1993);
H1 Collaboration, I. Ahmed {\it et al.}, Z. Phys. {\bf C64}, 545 (1994).
H1 Collaboration, I. Aid {\it et al.}, Z. Phys. {\bf C71}, 211 (1996).

\bibitem{ZEUS}
ZEUS Collaboration, M. Derrick {\it et al.}, Phys. Lett. {\bf B306}, 
173 (1993).

\bibitem{D0}
D0 Collaboration, S. Abachi {\it et al.}, Phys. Rev. Lett. {\bf 72}, 
965 (1994).

\bibitem{CDF}
CDF Collaboration, F. Abe {\it et al.}, Phys Rev. {\bf D48}, 3939 (1993).

\bibitem{ALEPH}
ALEPH Collaboration, DeCamp {\it et al.}, Phys. Rept. {\bf 216}, 
253 (1992).

\bibitem{DELPHI}
DELPHI Collaboration, P. Abreu {\it et al.}, Phys. Lett. {\bf B316}, 
620 (1993).

\bibitem{L3}
L3 Collaboration, Adriani {\it et al.}, Phys. Rept. {\bf 236}, 
1 (1993).

\bibitem{OPAL}
OPAL Collaboration, G. Alexander {\it et al.}, Phys. Lett. {\bf B263}, 
123 (1991).

\bibitem{h-r} For a comprehensive review see
J.L. Hewett and T.G. Rizzo, Phys. Rept. {\bf 183}, 193 (1989).
See also T.G. Rizzo, Phys. Lett. {\bf 192B}, 125 (1987);
H. Dreiner {\it et al.} Mod. Phys. Lett. {\bf A3}, 443 (1988).

\bibitem{DG3}
M. A. Doncheski and S. Godfrey, Phys. Lett.{\bf B},  (1997, in press)
[{\tt hep-ph/9608368}].

\bibitem{DG1}
M. A. Doncheski and S. Godfrey, Phys. Rev. {\bf D49}, 6220 (1994).

\bibitem{DG2}
M. A. Doncheski and S. Godfrey, Phys. Rev. {\bf D51}, 1040 (1995).

\bibitem{eboli} O.J. \'Eboli, E.M. Gregores, M.B. Magro, P.G. Mercadante, 
and S.F. Novaes, Phys. Lett. {\bf B311}, 147 (1993).

\bibitem{leptoLN}
H. Nadeau and D. London, Phys. Rev. {\bf D47}, 3742 (1993);
G. B\'elanger, D. London and H. Nadeau, Phys. Rev. {\bf D49}, 3140 (1994).

\bibitem{HP}
J.L. Hewett and S. Pakvasa, Phys. Lett. {\bf B227}, 178 (1989).

\bibitem{misc}  For related papers see also:
J. E. Cieza Montalvo and O.J.P. \'Eboli, Phys. Rev. {\bf D47}, 837 (1993);
T.M. Aliev and Kh.A. Mustafaev, Yad. Fiz. {\bf 58}, 771 (1991);
V. Ilyin {\it et al.}, Phys. Lett. {\bf B351}, 504 (1995); 
erratum {\bf B352}, 500 (1995); Phys. Lett. {\bf B356}, 531 (1995).

\bibitem{opal-pn280}
OPAL Collaboration, OPAL physics note PN 280 (March, 1997).

\bibitem{buch}
W. Buchm\"uller, R. R\"uckl, and D. Wyler, Phys. Lett. {\bf B191}, 
442 (1987).

\bibitem{nic}
A. Nicolaidis, Nucl. Phys. {\bf B163}, 156 (1980).

\bibitem{DO}
D.W. Duke and J.F. Owens, Phys. Rev. {\bf D26}, 1600 (1982).

\bibitem{DG}
M. Drees and K. Grassie, Z. Phys. {\bf C28}, 451 (1985);  
M. Drees and R. Godbole, Nucl. Phys. {\bf B339}, 355 (1990).

\bibitem{GRV}
M. Gl\"uck, E. Reya and A. Vogt, Phys. Lett. {\bf B222}, 149 (1989); 
Phys. Rev. {\bf D45}, 3986 (1992); Phys. Rev. {\bf D46}, 1973 (1992).

\bibitem{LAC}
H. Abramowicz, K. Charchula, and A. Levy, Phys. Lett. {\bf B269}, 458 
(1991).

%\bibitem{gv}
%M. Gl\"uck and W. Vogelsang, Z. Phys. {\bf C57}, 309 (1993).

\bibitem{DGP}
Hadronic backgrounds in $e^+e^-$ and $e\gamma$ collisions and 
associated references are given in 
M.A. Doncheski, S. Godfrey, and K.A. Peterson, Phys. Rev. {\bf D55},
183 (1997) [{\tt hep-ph/9407348}].

\end{references}
\end{document}